\pdfoutput=1

\documentclass[10pt,twocolumn,letterpaper]{article}

\usepackage[pagenumbers]{cvpr}

\usepackage{cuted}

\usepackage{currfile}

\usepackage{caption}

\newcommand{\needcite}[1]{\textcolor{magenta}{[cite]}}

\usepackage{microtype}
\renewcommand{\paragraph}[1]{\vspace{.5em}\noindent\textbf{#1}}

\newcommand{\dB}{\,dB\xspace}
\newcommand{\m}{\ensuremath{-}}

\definecolor{cvprblue}{rgb}{0.21,0.49,0.74}
\usepackage[pagebackref,breaklinks,colorlinks,allcolors=cvprblue]{hyperref}
\hypersetup{pdftitle={Who Plays What: Same-Instrument Music Separation from Video},pdfauthor={Abhinaba Roy, Ambuj Mehrish, Sebastiano Vascon, Dorien Herremans}}

\def\confName{CVPR}
\def\confYear{2027}

\title{Who Plays What: Same-Instrument Music Separation from Video}

\author{Abhinaba Roy$^{1}$, Ambuj Mehrish$^{2}$, Sebastiano Vascon$^{2}$, Dorien Herremans$^{1}$\\[0.4em]
{\normalsize $^{1}$Singapore University of Technology and Design\qquad $^{2}$University of Venice}\\[0.8em]
{\small \textbf{Demo page:}~\url{https://amaai-lab.github.io/who_plays_what_demo/}}\\
{\small \textbf{Code:}~\url{https://github.com/AMAAI-Lab/who_plays_what}}\\
{\small \textbf{Dataset:}~\url{https://huggingface.co/datasets/amaai-lab/who_plays_what}}
}

\begin{document}
\maketitle

\begin{strip}
\centering
\includegraphics[width=0.9\linewidth]{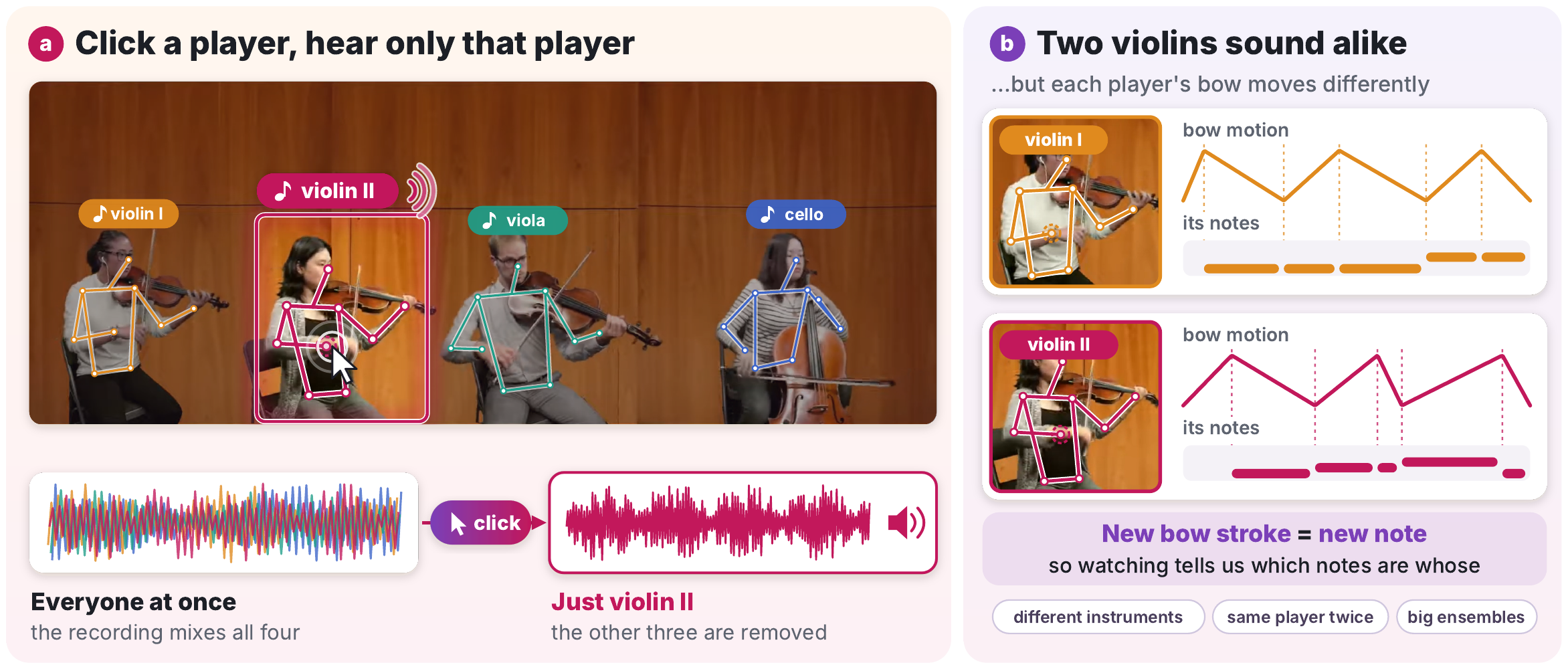}
\captionsetup{hypcap=false}
\captionof{figure}{\textbf{Click to separate.} Click a performer to hear only them; their bow motion tells apart even identical instruments.}
\label{fig:teaser}
\captionsetup{hypcap=true}
\end{strip}

\begin{abstract}
When two musicians in an ensemble play the same instrument, they sound alike and their instruments look alike. Audio-only separators, and audio-visual separators that rely on appearance, cannot tell them apart. What still differs is how each player moves, and when. We study same-instrument per-player separation: a user clicks one musician in a video and receives that musician's sound alone, even when other players in the video play the same instrument. Our model follows the clicked player and describes them with high level video features and the velocities of body and hand keypoints. A pretrained music separator attends to these tokens through cross-attention. Earlier motion-based separators were scored only on synthetic mixtures of solo videos. We give the first per-player evaluation of same-instrument separation on real ensemble recordings. On 58 players in URMP, each sharing an instrument with at least one other player in the piece, our model reaches a median SDR of 5.3\dB, while four released separators, audio-only and audio-visual, stay at or below 0.3\dB. The gap is wider on same-instrument duets, and our model also beats the audio-visual baselines on their own MUSIC benchmark. Tests show that the gain comes from movement: replacing the video with a still frame removes it, and so does shifting the video by a fraction of a second against the audio.
\end{abstract}

\section{Introduction}
\label{sec:intro}

Take a video of a string quartet, and suppose you want to hear only the second violin. A music separator~\cite{lu2024music,solovyev2026musicsourceseparationtrainingmsstunifiedframework} will give you a ``violin'' track, and that track contains both violins. An audio-visual separator can look at the video, but most of these models link how an instrument looks to how it sounds~\cite{zhao2019sound,chen2023iquery,huang2026high}, and the two violins look the same. When two players share an instrument, neither the timbre nor the appearance says which notes belong to whom.

One thing still differs between the two players: what each of them does, and when. A bow changes direction as a note begins. The fingers of the left hand move when the pitch changes. A flautist breathes in before a phrase. These movements are tied in time to that player's notes and to nobody else's. Sound of Motions~\cite{zhao2019sound} and Music Gesture~\cite{gan2020music} used this cue, with motion trajectories and with body and finger keypoints, and separated duets of the same instrument. Both, however, report SDR only on synthetic mixtures of solo videos, and we found no public code for either. To our knowledge, no separator has been scored player by player on real ensemble recordings in which two players share an instrument.

We study the problem in a form that a person can use directly (\cref{fig:teaser}).
A user clicks on one musician in the video and receives that musician's sound alone.
The click works like a promptable segmentation~\cite{ravi2025sam}: it says \emph{which} player is meant, not \emph{what kind} of instrument they play.
Possible uses are practice and teaching, where a student wants to hear their own part, and the remixing of concert videos, where a separated part can then be restored or remastered~\cite{melechovsky2025sonicmaster}.
Per-player tracks could also feed music analysis, such as similarity models that already train on separated stems~\cite{roy2026merit}.

Our model is assembled from pretrained parts.
A promptable tracker follows the clicked player through the video.
From the player's crop we take two descriptions of what the player does: tokens from a frozen self-supervised video encoder, and the frame-to-frame velocities of body and hand keypoints.
A pretrained music separator then attends to these tokens, through a cross-attention layer added after each of its time layers.
The separator already knows how instruments sound, so the video only has to tell it which player to keep. We train only on synthetic mixtures of solo performance videos, never on ensemble recordings.
Three of our four kinds of training mixture add a second player of the same instrument.
In the hardest kind, the second player is the same performer in the same video at another moment, so appearance cannot tell the two apart.

We then give the first per-player evaluation of same-instrument separation on real ensemble recordings.
We click 58 players in URMP~\cite{li2018creating}, each in a piece where another player has the same instrument, and score the output against the clicked player's own track.
Our model reaches a median SDR of 5.30\dB.
Four released separators, three of them audio-visual, stay at or below 0.30\dB, below a trivial reference that only turns the recording down (0.82\dB).
They return the other player of the pair about as loudly as the clicked one, while our model keeps that player 23.65\dB below it.
On pure duets, where two players of the same instrument play alone, our model reaches 18.73\dB against at most 1.75\dB for the baselines.
This failure of the baselines is not a setup error: on the MUSIC benchmark that two of them were trained for, they reproduce their published scores, and our model is 6.9\dB above the better one.

A good score does not show why a model works, so we test the role of movement directly.
The output follows the click: the correct click scores higher than a click on the other player for 51 of the 58 players, while the audio-visual baselines are near chance.
Replacing the video with one repeated frame keeps how the player looks and removes how they move, and the gain disappears.
Shifting the video by 0.16\,s against the audio keeps both the look and the movement and breaks only their timing, and over all players this also removes the gain.
Finally, a variant that sees only keypoint velocities, and no pixels, reaches 4.15\dB over all players and 12.18\dB on pure duets, against 0.52 and 2.95\dB for a model without video.
Movement alone can tell two players of the same instrument apart, and over all players the gain depends on its timing.

In summary, we make three contributions.
\begin{itemize}[leftmargin=*,nosep]
  \item \textbf{Task and benchmark.} To our knowledge, the first per-player evaluation of same-instrument separation on real ensemble recordings (\cref{sec:setup}).
  \item \textbf{Model.} A pretrained music separator that attends to the clicked player's video, trained only on synthetic mixtures of solo videos (\cref{sec:method}).
  \item \textbf{Evidence that movement does the work.} Wrong-click, still-frame, time-shift and keypoints-only tests (\cref{sec:motion}).
\end{itemize}

\section{Related work}
\label{sec:related}

\paragraph{Audio-only music separation.}
Music separators such as BS-RoFormer~\cite{lu2024music} and MVSep Mega~\cite{solovyev2026musicsourceseparationtrainingmsstunifiedframework} split a recording into instrument classes.
Given a quartet, they return one track that holds both violins, because nothing in their input says which violin is wanted.
Blind separation with permutation-invariant training can split two parts of the same timbre~\cite{sarkar2022ensembleset}, but its outputs come in no fixed order, so it cannot answer a request for one particular player.

\paragraph{Audio-visual separation by appearance.}
Sound of Pixels~\cite{zhao2018sound} trains a separator by mixing the audio of two videos and asking the model to undo the mix with the help of the frames.
Later models keep this recipe.
iQuery~\cite{chen2023iquery} uses one query per instrument, DAVIS-Flow~\cite{huang2026high} is a generative separator, and SAM Audio~\cite{shi2025sam} is a general separator that accepts a visual mask as a prompt.
Their training mixtures mostly pair different instruments, so the look of the instrument is enough to pick the right sound.
Two players of the same instrument give these models the same visual evidence twice.

\paragraph{Audio-visual separation by movement.}
Before deep learning, Parekh \etal~\cite{parekh2017motion,parekh2017guiding} tied the bow velocity of string players to the activations of a matrix factorisation, and Li \etal~\cite{li2017see,li2019online} matched the players in a video to the parts of a known score by relating bow strokes and vibrato to note events.
Among learned models, Sound of Motions~\cite{zhao2019sound} adds motion trajectories to the separator, and Music Gesture~\cite{gan2020music} adds body and finger keypoints and reports same-instrument separation for piano, flute and trumpet duets.
Body-part motion is likewise a standard cue for recognising human actions in video~\cite{roy2018discriminative}.
Both report SDR scores on synthetic mixtures of solo videos.
Slizovskaia \etal~\cite{slizovskaia2021conditioned} do test on real URMP recordings, but they treat two players of the same instrument as one source.
Our method needs one click and no score at test time.

\paragraph{Promptable models and frozen features.}
Our click plays the role of the point prompt in promptable segmentation, and our model builds on four frozen models.
SAM~2.1~\cite{ravi2025sam} turns a click into a mask and tracks it through the video.
V-JEPA~2~\cite{assran2025v}, a self-supervised video encoder, and RTMW~\cite{jiang2024rtmw}, a whole-body keypoint estimator, describe the tracked player.
SAM~3~\cite{carion2026sam} also accepts text, and we use it only to test a text prompt in place of the click.

\section{Method}
\label{sec:method}
\begin{figure*}[t]
    \centering
    \includegraphics[width=0.95\linewidth]{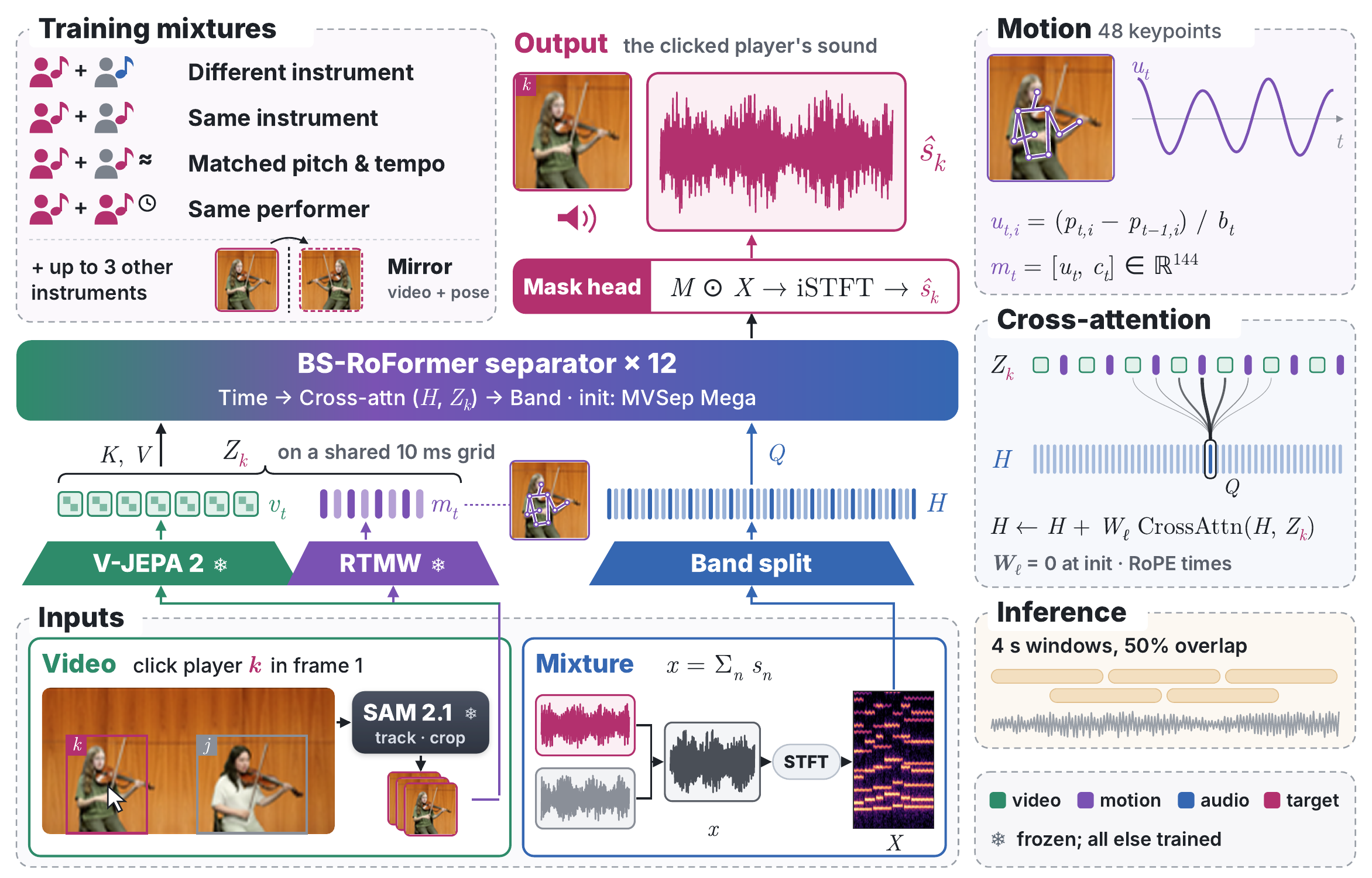}
    \caption{\textbf{Method overview.} A click selects the target player. Tracked crops provide video features and keypoint velocities, which condition a pretrained music separator through cross-attention. A mask head reconstructs the selected player's sound.}
    \label{fig:method}
\end{figure*}
\cref{fig:method} gives an overview of our method. Let $x\in\mathbb{R}^{T}$ be an audio recording containing $N$ players, with
\begin{equation}
    x=\sum_{n=1}^{N}s_n,
    \qquad
    \hat{s}_k=f_{\theta}(x,Z_k),
    \label{eq:task}
\end{equation}
where $s_n$ is the sound of player $n$ and $Z_k$ represents the player selected by a click in the first video frame. The objective is to recover $s_k$, even when another player $j\neq k$ plays the same instrument, so that instrument identity alone cannot select the target.

\subsection{Player Representation}
\label{sec:tokens}

A frozen SAM~2.1~\cite{ravi2025sam} with the Hiera-L backbone segments the clicked player and tracks the mask across the video. We take a square crop around the mask in each frame and encode crops with two frozen models: the self-supervised video encoder V-JEPA~2~\cite{assran2025v} for visual features and the whole-body keypoint estimator RTMW~\cite{jiang2024rtmw} for body and hand motion.

V-JEPA~2 ViT-L produces a $4\times4$ grid of 1024-dimensional tokens for every two frames, yielding 16 spatial tokens at 12.5 time steps per second. These features retain information about the player, instrument, and visible performance.

RTMW detects 48 body and hand keypoints per frame. For position $p_{t,i}\in\mathbb{R}^{2}$ and confidence $c_{t,i}$ of keypoint $i$, we compute a box-normalised velocity and concatenate the velocities and confidences:
\begin{equation}
\begin{aligned}
    u_{t,i}
    &=\frac{p_{t,i}-p_{t-1,i}}{b_t},\\
    m_t
    &=[u_{t,1},\ldots,u_{t,48},
       c_{t,1},\ldots,c_{t,48}]
       \in\mathbb{R}^{144},
\end{aligned}
\label{eq:velocity}
\end{equation}
where $b_t$ is the player's bounding-box size in frame $t$. We obtain one motion vector per frame at 25\,Hz. Using velocities rather than absolute positions emphasises movement and removes direct dependence on the player's location.

The video and keypoint tokens together form $Z_k$. Because these tokens and the separator's audio tokens have different sampling rates, we assign them timestamps on a shared 10\,ms grid. Rotary position embeddings~\cite{su2024roformer} encode these timestamps, allowing cross-attention to relate a player's movement to the corresponding sound.

\subsection{Player-Conditioned Separator}
\label{sec:separator}

We initialise the audio backbone from MVSep Mega~\cite{solovyev2026musicsourceseparationtrainingmsstunifiedframework}, a BS-RoFormer~\cite{lu2024music} pretrained to separate 53 instrument stems. The model computes an STFT with a 2048-sample window and a 512-sample hop, divides the spectrum into bands, and processes the band features through 12 blocks of width 256. Block $\ell$ applies a temporal transformer $\mathrm{Time}_{\ell}$, then a band-wise transformer $\mathrm{Band}_{\ell}$. We retain the pretrained band-split module and all 12 blocks, comprising 26.3M parameters.

We add a cross-attention layer after each $\mathrm{Time}_{\ell}$, with the audio features $H$ as queries and the player tokens $Z_k$ as keys and values:
\begin{align}
    H &\leftarrow \mathrm{Time}_{\ell}(H), \nonumber\\
    H &\leftarrow H+
    W_{\ell}\,\mathrm{CrossAttn}_{\ell}(H,Z_k),
    \label{eq:block_2}\\
    H &\leftarrow \mathrm{Band}_{\ell}(H). \nonumber
\end{align}
This places player-specific information within the separator's temporal processing, before information is exchanged across frequency bands. The output projection $W_{\ell}$ is initialised to zero, preserving the pretrained backbone's behaviour at initialisation.

We replace Mega's 53 instrument-specific output heads with a single head that predicts a complex mask $M$ for the selected player, so that $\hat{s}_k=\mathrm{iSTFT}(M\odot X)$ with $X=\mathrm{STFT}(x)$. We train the backbone, cross-attention layers, and the mask head, 43.9M parameters in total.

\subsection{Training}
\label{sec:training}

We train on synthetic mixtures constructed from solo performance videos in Solos~\cite{montesinos2020solos} and MUSIC-21~\cite{zhao2019sound,zhao2018sound}. The data cover 26 instruments and contain 1,088 training, 51 validation, and 78 test recordings. We exclude the 158 recordings belonging to the MUSIC and MUSIC-21 benchmark test sets, and we do not train on URMP.

Each example combines a four-second target clip with interfering audio. Given target audio $s_a$, its player tokens $Z_a$, and interference $s_b$, the model predicts
\begin{equation}
    x=s_a+s_b,
    \qquad
    \hat{s}_a=f_{\theta}(x,Z_a).
    \label{eq:training_mixture}
\end{equation}
Only the target video's tokens are provided, so the separator must use them to identify the corresponding source in the mixture.

The interfering audio contains a second player, drawn in one of four ways.
The second player plays a different instrument (probability 0.20), the same instrument in another video (0.30) or the same instrument in another video matched to the target clip in pitch and tempo (0.25), or is the same performer at another time in the target's video, with the same instrument and recording environment (0.25).
All four cases are mixed from the first training step, and we also add up to three clips of other instruments.
The same-instrument and same-performer cases make instrument identity and appearance unreliable shortcuts for source selection, which pushes the model toward the timing of the player's movement.

To account for viewing direction, we mirror player videos horizontally with probability $0.5$. V-JEPA~2 then sees the mirrored frames, and for the keypoints we negate the horizontal velocities and swap left and right.

We optimise an $\ell_1$ waveform loss $\|\hat{s}_a-s_a\|_1$ together with complex-spectrogram $\ell_1$ losses $\|\mathrm{STFT}_r(\hat{s}_a)-\mathrm{STFT}_r(s_a)\|_1$ at five resolutions $r$.

Training uses AdamW with batches of 32 four-second mixtures. The pretrained backbone receives one sixth of the learning rate used for the new layers. We select the checkpoint with the highest validation target-source SNR. At inference, recordings are processed in four-second windows with 50\% overlap. The supplementary material gives the loss weighting, the resolutions and further optimisation details.

\section{Experimental setup}
\label{sec:setup}

\begin{table*}[t]
  \centering
  \footnotesize
  \setlength{\tabcolsep}{3.5pt}
  \begin{tabular}{@{}llcccccccc@{}}
    \toprule
    & & \multicolumn{6}{c}{URMP, real ensemble recordings} & \multicolumn{2}{c}{Held-out mixtures} \\
    \cmidrule(lr){3-8} \cmidrule(l){9-10}
    & & \multicolumn{2}{c}{All (58)} & Strings & Winds \& & Pure & With other & All & Pure Duets $^{\ddagger}$ \\
    \cmidrule(lr){3-4}
    Method & Input & Median [95\% interval] & Ours better & (28) & brass (30) & duets (8) & instr.\ (50) & (500) & only (134) \\
    \midrule
    Unchanged recording          & none  & \m6.29 [\m6.79, \m5.61] & 58 & \m6.71 & \m5.77 & 0.00   & \m6.58 & \m4.46 & \m0.26 \\
    Scaled recording, $x/N$      & none  & 0.82 [0.66, 1.00]       & 46 & 0.73   & 1.02   & 2.91   & 0.80   & 1.09   & 2.65 \\
    \midrule
    SAM Audio                    & A + V & \m3.92 [\m4.58, \m3.16] & 57 & \m3.97 & \m3.73 & \m0.20 & \m4.05 & \m2.47 & \m0.07 \\
    iQuery                       & A + V & \m0.45 [\m0.78, \m0.10] & 51 & \m0.62 & \m0.12 & 0.29   & \m0.49 & \m0.66$^{\ast}$ & 0.46$^{\ast}$ \\
    DAVIS-Flow                   & A + V & 0.01 [\m0.09, 0.01]     & 47 & 0.01   & 0.01   & 0.02   & 0.00   & \m0.18 & 0.29 \\
    MVSep Mega                   & A     & 0.30 [0.19, 0.78]       & 42 & 0.43   & 0.30   & 1.75   & 0.28   & 0.26$^{\dagger}$ & 0.14$^{\dagger}$ \\
    \midrule
    \textbf{Ours}                & A + V & \textbf{5.30} [1.86, 10.31] &  & \textbf{6.20} & \textbf{3.62} & \textbf{18.73} & \textbf{3.02} & \textbf{9.56} & \textbf{14.36} \\
    \bottomrule
  \end{tabular}
  \caption{\textbf{Main results.}
  Median SDR in dB, higher is better. A: audio, V: video.
  \emph{Ours better}: players on which our model beats the row.
  \emph{Pure duets}: pieces with only the two players of the pair.
  $^{\ast}$Only mixtures of iQuery's 11 classes (220, 63 with two players).
  $^{\dagger}$Only 408 mixtures (114) for which Mega has a stem. $^{\ddagger}$ Two players playing the same instrument.}
  \label{tab:main}
\end{table*}

\paragraph{URMP same-instrument test set.}
URMP~\cite{li2018creating} contains chamber music pieces with video and a separate audio track for each player.
Its parts were recorded one player at a time and then assembled into one recording and one video.
We take the 29 pairs of players who share an instrument within a piece, which gives 58 clicked players: 26 violins, 2 violas, 12 flutes, 16 trumpets and 2 saxophones.
The pieces have two to five players.
Eight of the 58 players are in \emph{pure duets}, four pieces in which the pair plays alone, and the other 50 play alongside other instruments.
Each click is three points on the player's body (nose, shoulders, hips), placed by a person detector on the first frame and checked by eye.
\cref{sec:prompts} tests other prompts.

\paragraph{Metric and statistics.}
We score each output with the signal-to-distortion ratio (SDR) of museval~\cite{stoter20182018} against the clicked player's own track, at URMP's sampling rate of 48\,kHz, to which all outputs are resampled.
A player's score is the median over 1\,s windows, and we report the median over players.
For the median over all 58 players we give a 95\% bootstrap interval from 10,000 resamples, keeping the two players of a pair together since they come from the same piece.

\paragraph{Two trivial references.}
Returning the recording at a lower volume raises SDR without separating anything.
We therefore report the \emph{unchanged recording} $x$ and the \emph{scaled recording} $x/N$, where $N$ is the number of players.
A method counts as working only if it beats the scaled recording.

\paragraph{Held-out synthetic mixtures.}
As a second test, we build 500 mixtures of 4\,s from the 78 held-out recordings of our Solos and MUSIC-21 split (\cref{sec:training}).
Each mixture holds two players of one instrument, and 366 of them also hold one to three other instruments.
The second player comes from another video (170 mixtures), from another video matched in pitch and tempo (149), or from the same video at another moment (181).
No design decision or checkpoint choice used these mixtures.
Since they are built like our training mixtures, they favour our model over the baselines.

\paragraph{Baselines.}
We compare with four released models, each run as its authors released it and not retrained on our data.
\textbf{SAM Audio}~\cite{shi2025sam} receives the SAM~2.1 mask of the clicked player as its visual prompt, \textbf{iQuery}~\cite{chen2023iquery} receives the player's crop with the instrument name as its query, and \textbf{DAVIS-Flow}~\cite{huang2026high} receives the player's crop.
\textbf{MVSep Mega}~\cite{solovyev2026musicsourceseparationtrainingmsstunifiedframework} is audio-only.
Given the instrument, it returns that instrument's stem, which holds both players of the pair.
iQuery and DAVIS-Flow output audio at 11\,kHz, so they lose everything above 5.5\,kHz in our 48\,kHz scoring.
We did not find any public code for Sound of Motions and Music Gesture.

\paragraph{Ablation models.}
We train ablations on our training data.
The \emph{curriculum model} changes only the order of the training mixtures (\cref{sec:training}): 4k steps on each of three cases in turn (a different instrument, the same instrument from another video, the same performer at another moment), then the full mixtures.
All other ablations mix all the cases, like our model.
The \emph{no-video model} has no player tokens and no cross-attention.
The \emph{keypoints-only model} drops the V-JEPA~2 tokens, so it sees how the player moves but not a single pixel.
The \emph{no-mirror model} is trained without mirror augmentation, and a last model starts the no-mirror model's backbone from random weights instead of Mega's.

\section{Results}
\label{sec:results}

\subsection{Comparison with released separators}
\label{sec:main}

No released baseline beats the scaled recording in any URMP column of \cref{tab:main}.
The best of them is the audio-only Mega, at 0.30\dB over all players, and the three audio-visual baselines stay at or below 0.01\dB.
Our model reaches 5.30\dB over all 58 players, and its 95\% interval, from 1.86 to 10.31\dB, lies above those of every baseline and both trivial references.
The interval is wide because the method works very well on some pieces and fails on others (\cref{sec:failures}).
On the 8 players in pure duets it reaches 18.73\dB and is ahead of every baseline for all 8.
The held-out synthetic mixtures give the same picture with larger margins: 9.56\dB for our model against at most 0.46\dB for the baselines.

To separate the effect of the video from that of our training data, we train the same model without video on the same data (\cref{tab:ablation}, row c).
It reaches 0.52\dB over all players and 2.95\dB on pure duets, close to the scaled recording, as expected from a model that cannot know which player is wanted.
The gain comes from the video, not from the training data.

\begin{table}[h]
  \centering
  \footnotesize
  \setlength{\tabcolsep}{5pt}
  \begin{tabular}{@{}lccc@{}}
    \toprule
    & \multicolumn{2}{c}{Partner} & Other instruments \\
    \cmidrule(lr){2-3}
    Method & All (58) & Pure duets (8) & (50) \\
    \midrule
    SAM Audio                        & \m0.61 & 0.14   & \m17.39 \\
    iQuery                           & 0.08   & \m0.76 & 2.32 \\
    DAVIS-Flow                       & 0.26   & 1.07   & 4.78 \\
    MVSep Mega                       & 0.00   & 0.00   & 18.09 \\
    \midrule
    No-video model$^{\ddagger}$      & 0.00   & 0.00   & \m0.82 \\
    Curriculum model$^{\ddagger}$    & 21.14  & 32.81  & 14.23 \\
    \textbf{Ours}                    & \textbf{23.65} & \textbf{39.01} & \textbf{21.23} \\
    \bottomrule
  \end{tabular}
  \caption{\textbf{What is left in the output.}
  Level of each unwanted sound below the clicked player, in dB; higher means less leakage.
  \emph{Partner}: the other player of the same-instrument pair.
  $^{\ddagger}$Ablations.}
  \label{tab:leak}
\end{table}

\paragraph{What the baselines leave in.}
SDR counts all unwanted sound together. To see which sound remains (\cref{tab:leak}), we fit each output as a weighted sum of three true signals: the clicked player, the other player of the pair, which we call the partner, and the other instruments.
This is a simple gain fit, not the SIR of museval (supp. material).
Every baseline, like our no-video model, keeps the partner about as loud as the clicked player, between \m0.76 and 1.07\dB, so none of them splits two players of one instrument.
Mega removes the other instruments well (18.09\dB) but not the partner (0.00\dB): it separates instruments, not players.
Our model keeps the partner 23.65\dB below the clicked player, 39.01\dB in pure duets, and the other instruments 21.23\dB below.

\begin{table}[t]
  \centering
  \footnotesize
  \setlength{\tabcolsep}{6pt}
  \begin{tabular}{@{}lcc@{}}
    \toprule
    Method & Mean SDR [95\% interval] & In its paper \\
    \midrule
    The mixture itself           & 0.16                  & \\
    iQuery                       & 11.61 [10.88, 12.36]  & 11.17 \\
    DAVIS-Flow                   & 12.21 [11.40, 13.03]  & 12.01 \\
    \textbf{Ours}                & \textbf{19.08} [18.22, 19.97] & \\
    \bottomrule
  \end{tabular}
  \caption{\textbf{The baselines' own benchmark.}
  200 mixtures of two different instruments from the MUSIC and MUSIC-21 test recordings, scored with the baselines' protocols.}
  \label{tab:music}
\end{table}

\paragraph{The baselines' own benchmark.}
We also score iQuery, DAVIS-Flow and our model on the MUSIC benchmark that these two baselines were trained and tested on (\cref{tab:music}).
We build 200 mixtures of two different instruments from 77 test recordings of MUSIC and MUSIC-21, none of which was used to train our model or to pick its checkpoint.
Both baselines reproduce their published scores within 0.5\dB, so their weak results on URMP are unlikely a setup error.
Our model is 6.9\dB above the better baseline and better on at least 183 of the 200 mixtures.
Every mixture here holds two different instruments, so this benchmark tests the usual task and not the same-instrument case.
Our model also starts from Mega, which separates 53 kinds of stems and has therefore seen far more instruments than the 11 of MUSIC.

\subsection{Listening test}
\label{sec:listening}

SDR compares waveforms and depends on loudness, so we also ask listeners, in a MUSHRA test~\cite{schoeffler2015towards}.
In each trial the listener sees a frame of the video with the clicked player marked and hears that player's true track as the reference.
The listener then rates seven unlabelled clips from 0 to 100 for how close each one is to the reference: the hidden reference, our model, the four baselines, and the unchanged recording as the anchor.

There are 10 trials, each from a different piece, picked at random: 3 duets, 2 trios, 3 quartets and 2 quintets.
The excerpts of 10\,s are chosen by a fixed rule before anyone listened to an output, and all clips are normalised to \m23\,LUFS, so loudness cannot decide the result.
16 listeners took part. 
Our model scores 69.0 over all trials and 88.0 on the pure duets (\cref{tab:mushra}). The best baselines score 30.5 (SAM Audio) and 30.2 (MVSep Mega) over all trials, and 38.5 (SAM Audio) on the pure duets.
Every one of the 16 listeners gives our model a higher mean rating than every baseline ($p = 1.2 \times 10^{-4}$, exact Wilcoxon signed-rank test over the listeners' mean ratings, Holm-corrected).
The baselines barely improve on the unchanged recording: over all trials, iQuery and DAVIS-Flow score below it, and SAM Audio and MVSep Mega are less than 5 points above.
\begin{table}[!t]
  \centering
  \footnotesize
  \setlength{\tabcolsep}{5pt}
  \begin{tabular}{@{}lcc@{}}
    \toprule
    Method & All trials (10) & Pure duets (3) \\
    \midrule
    Hidden reference              & 95.3 (91.7, 98.1) & 95.5 (91.4, 98.6) \\
    \midrule
    SAM Audio                     & 30.5 (16.8, 45.2) & 38.5 (25.1, 52.4) \\
    iQuery                        & 22.1 (14.2, 30.3) & 19.7 (11.4, 29.6) \\
    DAVIS-Flow                    & 18.2 (10.7, 26.8) & 11.4 (1.4, 27.2) \\
    MVSep Mega                    & 30.2 (21.1, 40.6) & 29.8 (18.9, 40.5) \\
    \textbf{Ours}                 & \textbf{69.0} (50.0, 85.2) & \textbf{88.0} (73.8, 98.1) \\
    \midrule
    Anchor (unchanged recording)  & 26.0 (16.3, 36.6) & 31.3 (19.0, 45.4) \\
    \bottomrule
  \end{tabular}
  \caption{\textbf{Listening test (MUSHRA).}
  Mean score from 0 to 100 over 16 listeners, higher is better, with the 95\% bootstrap interval over listeners and trials in brackets. }
  \label{tab:mushra}
\end{table}

\begin{figure}[t]
  \centering
  \includegraphics[width=\linewidth]{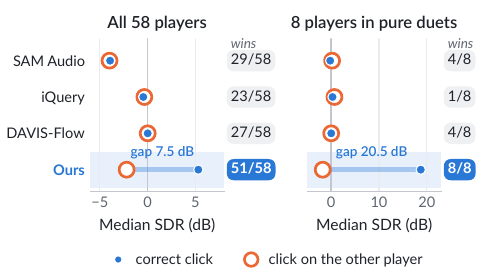}
  \caption{\textbf{Does the output follow the click?}
  Score against a player’s track with the click on that player and on the other player of the pair.
  \emph{Wins}: players for whom the correct click scores higher. Over all players, the baselines win about half the time.}
  \label{fig:click}
\end{figure}

\subsection{Does the output follow the click?}
\label{sec:click}

A promptable separator must return a different sound when a different player is clicked.
We test this for every player by clicking the \emph{other} player of the pair and scoring the output against the first player's track.
If a method follows the click, the correct click scores higher.

For our model the correct click scores higher for 51 of the 58 players and for all 8 players in pure duets (\cref{fig:click}).
A wrong click drops the median from 5.30 to \m2.19\dB over all players and from 18.73 to \m1.76\dB on pure duets.
The three audio-visual baselines win for 23 to 29 of the 58 players.
SAM Audio, for example, mostly returns the same, more prominent player whichever player is clicked.

\begin{figure}[t]
  \centering
  \includegraphics[width=\linewidth]{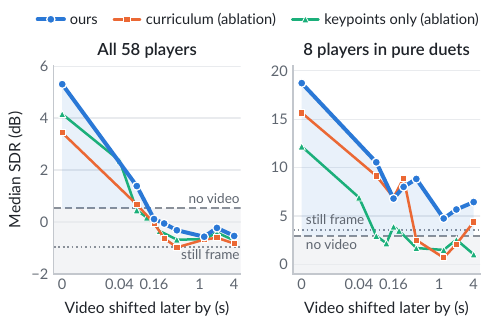}
  \caption{\textbf{Movement, with time, carries the separation.}
  Median SDR as the video is delayed against the audio (log axis above 0.04\,s).
  Over all players, a shift of 0.08 to 0.16\,s brings all models to the no-video model (dashed line).
  In pure duets our model keeps part of its gain (blue area).
  Dotted: our model with a still frame.}
  \label{fig:motion}
\end{figure}

\subsection{Does the model use movement?}
\label{sec:motion}

The click test does not show whether the model recognises \emph{who} the player is from the crop, by position or clothing, or uses \emph{what the player does}.
We test this in three ways, from the weakest test to the strongest (\cref{fig:motion}).

\paragraph{Still frame.}
We replace each player's video by its middle frame, repeated, and set the keypoint velocities to zero.
The model still sees who the player is and what they play, but not how they move.
The median falls from 5.30 to \m0.97\dB over all players and from 18.73 to 3.56\dB on pure duets.
The correct click now wins for only 26 of the 58 players, which is chance, and in the duets the model mostly loses one of the two players.
One caveat is that the model never saw a motionless player in training, so a still frame is also an unusual input, which may be why it scores below the no-video model (0.52\dB).

\paragraph{Time shift.}
We keep the real video and shift it later than the audio, wrapping around at the end of the piece.
The player looks and moves as before, and only the timing between movement and sound changes, so a model that only recognised the player would keep its score.
We test our model and two ablations, the curriculum model and the keypoints-only model, without retraining.
Models with V-JEPA~2 tokens can be shifted in steps of 0.08\,s, the time one token covers, and the keypoints-only model by a single video frame, 0.04\,s.

Over all players, the gain needs the movement to line up with the sound to within about 0.1\,s.
A shift of 0.08\,s takes our model from 5.30 to 1.38\dB.
At 0.16\,s all three models are below the no-video model, and larger shifts, up to 4\,s, do not bring the gain back.
For the keypoints-only model one frame already matters: a shift of 0.04\,s takes its duet score from 12.18 to 6.90\dB.
In pure duets our model keeps part of its gain under every shift, between 4.73 and 10.54\dB, against 18.73\dB in sync and 2.95\dB without video.
There it can partly tell the two players apart from something that does not depend on timing, such as who the player is.
The keypoints-only model, which sees no appearance, falls to around the no-video level from 0.4\,s on. So in duets, timing is most but not all of what our model uses.

\paragraph{Keypoints only.}
The keypoints-only model never sees a pixel: it receives only the velocities and detection confidences of 48 keypoints, with no absolute positions in the frame.
Because it is trained this way, the objection to the still frame does not apply.
It reaches 12.18\dB on pure duets and 4.15\dB over all 58 players, against 2.95 and 0.52\dB for the no-video model (\cref{tab:ablation}).
Its correct click wins for 51 of the 58 players, as for our model.
Keypoint motion alone separates two players of the same instrument on real recordings. The three tests agree. Over all players, taking away the movement or its timing removes the gain, and keeping only the movement keeps most of it.

\paragraph{Same performer, another moment.}
The held-out mixtures point the same way.
In 181 of them the second player is the same performer in the same video at another moment, so the two look identical.
The no-video model reaches 1.39\dB, the keypoints-only model 3.86\dB and our model 6.31\dB.

\paragraph{Timing, not amount.}
Players who move more are not separated better.
The gain of our model over the no-video model does not correlate with a player's mean keypoint speed (Spearman 0.07 for the arms and 0.03 for the hands, $p \ge 0.6$).
Mean speed measures how much a player moves, not whether the movement lines up with the notes.

\begin{table}[!t]
  \centering
  \footnotesize
  \setlength{\tabcolsep}{3.4pt}
  \begin{tabular}{@{}l@{\hspace{3pt}}lccccc@{}}
    \toprule
    & & \multicolumn{4}{c}{URMP} & Held- \\
    \cmidrule(lr){3-6}
    & Model & All & Str. & W\&B & Duets & out \\
    \midrule
    (a) & \textbf{Ours}                     & \textbf{5.30} & \textbf{6.20} & 3.62 & \textbf{18.73} & 9.56 \\
    (b) & curriculum                        & 3.43 & 3.44 & 2.73   & 15.66 & \textbf{9.90} \\
    \midrule
    (c) & (a), no video                     & 0.52 & 0.61 & 0.38   & 2.95  & 1.24 \\
    (d) & (a), keypoints only               & 4.15 & 5.04 & \textbf{3.83} & 12.18 & 6.46 \\
    (e) & (a), no mirror augm.              & 3.29 & 5.19 & 0.66   & 11.04 & 9.54 \\
    (f) & (e), test video at 960 px         & 0.44 & 4.34 & \m0.24 & 6.75  & 9.54 \\
    (g) & (f), backbone not pretrained        & 0.40 & 0.71 & 0.36   & 0.51  & 2.08 \\
    \bottomrule
  \end{tabular}
  \caption{\textbf{Ablations.} Median SDR in dB. Rows (b) to (e) change one thing in (a), and rows (f) and (g) one thing in the row they name. Str.: strings. W\&B: winds and brass.}
  \label{tab:ablation}
\end{table}

\subsection{Ablations}
\label{sec:ablations}
In \cref{tab:ablation}, rows (b) to (e) change one thing in our model: the training schedule, the video input or the mirror augmentation.
Without any video (row c) the model falls to the level of the scaled recording and cannot follow a click at all.

\paragraph{Curriculum.}
Mixing the four cases of \cref{sec:training} from the first step (row a of \cref{tab:ablation}) is better on URMP than introducing them one after another (row b, \cref{sec:setup}): 5.30 against 3.43\dB over all players, higher for 44 of the 58 players, and 18.73 against 15.66\dB on pure duets.
On the held-out mixtures the two do not differ by much (9.56 against 9.90\dB).
The supplementary material compares them by group, by prompt and under time shift.

\paragraph{Video tokens.}
Row (d) drops the V-JEPA~2 tokens and keeps the keypoints.
Adding the video tokens raises the median from 4.15 to 5.30\dB over all players and from 12.18 to 18.73\dB on pure duets.
They also help on the same-performer mixtures, where appearance cannot help (3.86 to 6.31\dB), so they carry information about what the player does that the keypoints miss.
For winds and brass the two models are close (3.83 against 3.62\dB).
We have not trained a model with video tokens and no keypoints.

\paragraph{Mirroring and crop size.}
Mirror augmentation helps in every URMP column (rows e and a): winds and brass rise from 0.66 to 3.62\dB, pure duets from 11.04 to 18.73\dB, and the correct click wins for 51 instead of 47 of the 58 players.
The held-out score does not change (9.54 and 9.56\dB).
The video tokens are also sensitive to how large the player appears.
Our training crops are about 360\,px wide and URMP crops about 480\,px.
When the URMP video is downscaled to 960\,px, the crops shrink to about 240\,px and the no-mirror model drops from 3.29 to 0.44\dB over all players.

\paragraph{Pretrained separator.}
Without Mega's weights (row g against row f), the score on pure duets falls from 6.75 to 0.51\dB and the held-out score from 9.54 to 2.08\dB.
Scored at full resolution it also loses almost all of the gain (0.37 against 3.29\dB over all players for row e).
It still follows the click (48 of the 58 players at full resolution), but it separates poorly.
The pretrained backbone supplies the ability to separate, and the video supplies the choice of player.

\begin{table}[t]
  \centering
  \footnotesize
  \setlength{\tabcolsep}{3.4pt}
  \begin{tabular}{@{}lccccc@{}}
    \toprule
    & & & & Pure & Within 1\,dB \\
    Prompt & All & Str. & W\&B & duets & of default \\
    \midrule
    Three points (default)       & 5.30   & 6.20   & 3.62   & 18.73 & \\
    One point (nose)             & 4.25   & 4.25   & 4.24   & 19.18 & 48 of 58 \\
    Box                          & 4.64   & 5.39   & 4.58   & 18.94 & 49 of 58 \\
    Three points, moved          & 5.63   & 5.97   & 3.58   & 19.12 & 49 of 58 \\
    Text, via SAM~3              & \m0.77 & \m1.55 & \m0.77 & 18.73 & \\
    \bottomrule
  \end{tabular}
  \caption{\textbf{Other prompts.} Our model, median SDR in dB, no retraining.
  \emph{Moved}: each point is moved at random by up to 10\% of the player's height.}
  \label{tab:prompts}
\end{table}

\subsection{Other prompts}
\label{sec:prompts}

Besides the default three points (\cref{sec:setup}), we also try the nose alone, a box on the first frame, and the three points each moved at random, every time through the whole pipeline from SAM~2.1 tracking to the separator.
All of them work (\cref{tab:prompts}): 48 or 49 of the 58 players stay within 1\dB of the default prompt, and the pure duets stay between 18.94 and 19.18\dB.
The differences come from slightly different SAM~2.1 masks, not from a different player.

A text prompt can also replace the click.
SAM~3~\cite{carion2026sam} turns a phrase such as ``left violin'' into a mask, and the separator is unchanged.
SAM~3 picks the intended player for 33 of the 58 players, including all 8 in pure duets, where the result is the same as a click.
It picks another player for 11, for example, a viola for a violin, and finds no player for 14, most of them flutes, in which case nothing is separated.

\begin{figure}[t]
  \centering
  \includegraphics[width=\linewidth]{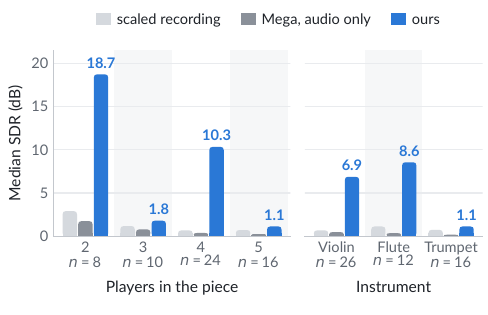}
  \caption{\textbf{Where the method works.}
  Median SDR by the number of players per piece (left) and by instrument (right); $n$: clicked players.
  Viola and saxophone, one pair each, are not drawn.}
  \label{fig:groups}
\end{figure}

\subsection{Where it works and where it fails}
\label{sec:failures}

Our model is ahead of every baseline and in every group of \cref{fig:groups}.
It is strongest in pure duets (18.73\dB) and in four-player pieces (10.34\dB), and for flutes and violins.
Three groups are weak: three-player pieces (1.80\dB), five-player pieces (1.14\dB) and trumpets (1.14\dB), each only a little above the scaled recording.
The trumpets fail where they play next to a horn, a trombone or a tuba, and the trumpet duet itself works well (19.73 and 20.58\dB).
Two violin pairs in five-player pieces also fail, although the tracking is correct and the two parts of each pair do not play the same notes, and we could not find the reason.
Overall our model is below the scaled recording for 12 of the 58 players: it helps most players a lot and some not at all.

\section{Conclusion and Limitations}
\label{sec:conclusion}
We studied how to separate a clicked musician when another player uses the same instrument. Trained only on synthetic mixtures, our video-conditioned separator achieves median SDRs of 5.30\dB across 58 URMP players and 18.73\dB on same-instrument duets. Motion controls show that the improvement depends on audio--visual alignment. Separation remains less effective in some larger ensembles and mixed-brass pieces. The evaluation uses separately recorded parts, some design choices were informed by URMP, and each setting was trained once. The model is trained on 26 instruments. The results may not hold for other instruments or other styles of music. Live ensembles and invalid-click rejection remain directions for further work, while possible overlap with MVSep Mega's undisclosed training data cannot be ruled out. 

\section*{LLM Usage Statement}
We used Claude (Anthropic), a large language model, as a writing and editing assistant. \textbf{Text:} The model edited author-written manuscript drafts and captions for clarity, concision, and consistent terminology and notation. \textbf{Figures:} The authors drew the diagrams manually using draw.io and later improved them with the model's assistance. The model wrote the plotting code for the results figures from our data. \textbf{Code}: Using Claude Code, the model wrote parts of the training, evaluation and ablation code. The authors reviewed and verified all model-edited text, figures, and code, and take full responsibility for the content of the paper.

{
    \small
    \bibliographystyle{ieeenat_fullname}
    \bibliography{references}

@String(CVPR= {IEEE Conf. Comput. Vis. Pattern Recog.})

@String(ICCV= {Int. Conf. Comput. Vis.})

@String(ECCV= {Eur. Conf. Comput. Vis.})

@String(ICASSP=	{ICASSP})

@String(ICLR = {Int. Conf. Learn. Represent.})

@String(CVPR  = {CVPR})

@String(ICCV  = {ICCV})

@String(ECCV  = {ECCV})

@String(ICLR  = {ICLR})

@inproceedings{lu2024music,
  title={Music source separation with band-split {RoPE} transformer},
  author={Lu, Wei-Tsung and Wang, Ju-Chiang and Kong, Qiuqiang and Hung, Yun-Ning},
  booktitle={ICASSP 2024-2024 IEEE International Conference on Acoustics, Speech and Signal Processing (ICASSP)},
  pages={481--485},
  year={2024},
  organization={IEEE}
}

@inproceedings{montesinos2020solos,
  title={{Solos}: A dataset for audio-visual music analysis},
  author={Montesinos, Juan F and Slizovskaia, Olga and Haro, Gloria},
  booktitle={2020 IEEE 22nd International Workshop on Multimedia Signal Processing (MMSP)},
  pages={1--6},
  year={2020},
  organization={IEEE}
}

@inproceedings{zhao2018sound,
  title={The sound of pixels},
  author={Zhao, Hang and Gan, Chuang and Rouditchenko, Andrew and Vondrick, Carl and McDermott, Josh and Torralba, Antonio},
  booktitle={European Conference on Computer Vision (ECCV)},
  pages={587--604},
  year={2018},
  organization={Springer}
}

@inproceedings{sarkar2022ensembleset,
  title={{EnsembleSet}: A new high quality synthesised dataset for chamber ensemble separation},
  author={Sarkar, Saurjya and Benetos, Emmanouil and Sandler, Mark and others},
  booktitle={ISMIR},
  pages={625--632},
  year={2022}
}

@misc{solovyev2026musicsourceseparationtrainingmsstunifiedframework,
  title={{Music-Source-Separation-Training (MSST)}: A Unified Framework for Training and Evaluating Music Demixing Models}, 
  author={Roman Solovyev and Ilya Kiselev and Alexander Stempkovskiy and Tatiana Gabruseva},
  year={2026},
  eprint={2607.23395},
  archivePrefix={arXiv},
  primaryClass={cs.SD},
  url={https://arxiv.org/abs/2607.23395}, 
}

@inproceedings{zhao2019sound,
  title={The sound of motions},
  author={Zhao, Hang and Gan, Chuang and Ma, Wei-Chiu and Torralba, Antonio},
  booktitle={2019 IEEE/CVF International Conference on Computer Vision (ICCV)},
  pages={1735--1744},
  year={2019},
  organization={IEEE}
}

@inproceedings{chen2023iquery,
  title={{iQuery}: Instruments as queries for audio-visual sound separation},
  author={Chen, Jiaben and Zhang, Renrui and Lian, Dongze and Yang, Jiaqi and Zeng, Ziyao and Shi, Jianbo},
  booktitle={2023 IEEE/CVF Conference on Computer Vision and Pattern Recognition (CVPR)},
  pages={14675--14686},
  year={2023},
  organization={IEEE}
}

@inproceedings{stoter20182018,
  title={The 2018 signal separation evaluation campaign},
  author={St{\"o}ter, Fabian-Robert and Liutkus, Antoine and Ito, Nobutaka},
  booktitle={International Conference on Latent Variable Analysis and Signal Separation},
  pages={293--305},
  year={2018},
  organization={Springer}
}

@article{li2018creating,
  title={Creating a multitrack classical music performance dataset for multimodal music analysis: Challenges, insights, and applications},
  author={Li, Bochen and Liu, Xinzhao and Dinesh, Karthik and Duan, Zhiyao and Sharma, Gaurav},
  journal={IEEE Transactions on Multimedia},
  volume={21},
  number={2},
  pages={522--535},
  year={2018},
  publisher={IEEE}
}

@article{su2024roformer,
  title={{RoFormer}: Enhanced transformer with rotary position embedding},
  author={Su, Jianlin and Ahmed, Murtadha and Lu, Yu and Pan, Shengfeng and Bo, Wen and Liu, Yunfeng},
  journal={Neurocomputing},
  volume={568},
  pages={127063},
  year={2024},
  publisher={Elsevier}
}

@inproceedings{carion2026sam,
  title={{SAM 3}: Segment anything with concepts},
  author={Carion, Nicolas and Gustafson, Laura and Hu, Yuan-Ting and Debnath, Shoubhik and Hu, Ronghang and Suris Coll-Vinent, Didac and Ryali, Chaitanya and Alwala, Kalyan Vasudev and Khedr, Haitham and Huang, Andrew and others},
  booktitle={International Conference on Learning Representations (ICLR)},
  year={2026}
}

@article{jiang2024rtmw,
  title={{RTMW}: Real-time multi-person {2D} and {3D} whole-body pose estimation},
  author={Jiang, Tao and Xie, Xinchen and Li, Yining},
  journal={arXiv preprint arXiv:2407.08634},
  year={2024}
}

@article{assran2025v,
  title={{V-JEPA 2}: Self-supervised video models enable understanding, prediction and planning},
  author={Assran, Mido and Bardes, Adrien and Fan, David and Garrido, Quentin and Howes, Russell and Muckley, Matthew and Rizvi, Ammar and Roberts, Claire and Sinha, Koustuv and Zholus, Artem and others},
  journal={arXiv preprint arXiv:2506.09985},
  year={2025}
}

@article{slizovskaia2021conditioned,
  title={Conditioned source separation for musical instrument performances},
  author={Slizovskaia, Olga and Haro, Gloria and G{\'o}mez, Emilia},
  journal={IEEE/ACM Transactions on Audio, Speech, and Language Processing},
  volume={29},
  pages={2083--2095},
  year={2021},
  publisher={IEEE}
}

@inproceedings{ravi2025sam,
  title={{SAM 2}: Segment anything in images and videos},
  author={Ravi, Nikhila and Gabeur, Valentin and Hu, Yuan-Ting and Hu, Ronghang and Ryali, Chaitanya and Ma, Tengyu and Khedr, Haitham and R{\"a}dle, Roman and Rolland, Chloe and Gustafson, Laura and others},
  booktitle={International Conference on Learning Representations (ICLR)},
  year={2025}
}

@inproceedings{gan2020music,
  title={Music gesture for visual sound separation},
  author={Gan, Chuang and Huang, Deng and Zhao, Hang and Tenenbaum, Joshua B and Torralba, Antonio},
  booktitle={2020 IEEE/CVF Conference on Computer Vision and Pattern Recognition (CVPR)},
  pages={10475--10484},
  year={2020},
  organization={IEEE}
}

@article{li2019online,
  title={Online audio-visual source association for chamber music performances},
  author={Li, Bochen and Dinesh, Karthik and Xu, Chenliang and Sharma, Gaurav and Duan, Zhiyao},
  journal={Transactions of the International Society for Music Information Retrieval},
  volume={2},
  number={1},
  pages={29--42},
  year={2019},
  publisher={Ubiquity Press}
}

@inproceedings{li2017see,
  title={See and listen: Score-informed association of sound tracks to players in chamber music performance videos},
  author={Li, Bochen and Dinesh, Karthik and Duan, Zhiyao and Sharma, Gaurav},
  booktitle={2017 IEEE International Conference on Acoustics, Speech and Signal Processing (ICASSP)},
  pages={2906--2910},
  year={2017},
  organization={IEEE}
}

@inproceedings{parekh2017guiding,
  title={Guiding audio source separation by video object information},
  author={Parekh, Sanjeel and Essid, Slim and Ozerov, Alexey and Duong, Ngoc QK and P{\'e}rez, Patrick and Richard, Ga{\"e}l},
  booktitle={2017 IEEE Workshop on Applications of Signal Processing to Audio and Acoustics (WASPAA)},
  pages={61--65},
  year={2017},
  organization={IEEE}
}

@inproceedings{parekh2017motion,
  title={Motion informed audio source separation},
  author={Parekh, Sanjeel and Essid, Slim and Ozerov, Alexey and Duong, Ngoc QK and P{\'e}rez, Patrick and Richard, Ga{\"e}l},
  booktitle={2017 IEEE International Conference on Acoustics, Speech and Signal Processing (ICASSP)},
  pages={6--10},
  year={2017},
  organization={IEEE}
}

@article{shi2025sam,
  title={{SAM Audio}: Segment anything in audio},
  author={Shi, Bowen and Tjandra, Andros and Hoffman, John and Wang, Helin and Wu, Yi-Chiao and Gao, Luya and Richter, Julius and Le, Matt and Vyas, Apoorv and Chen, Sanyuan and others},
  journal={arXiv preprint arXiv:2512.18099},
  year={2025}
}

@inproceedings{schoeffler2015towards,
  title={Towards the next generation of web-based experiments: A case study assessing basic audio quality following the {ITU-R} recommendation {BS.1534} ({MUSHRA})},
  author={Schoeffler, Michael and St{\"o}ter, Fabian-Robert and Edler, Bernd and Herre, J{\"u}rgen},
  booktitle={Proceedings of the 1st Web Audio Conference (WAC)},
  pages={1--6},
  year={2015}
}

@article{huang2026high,
  title={High-quality sound separation across diverse categories via visually-guided generative modeling},
  author={Huang, Chao and Liang, Susan and Tian, Yapeng and Kumar, Anurag and Xu, Chenliang},
  journal={International Journal of Computer Vision},
  volume={134},
  number={3},
  pages={104},
  year={2026},
  publisher={Springer}
}

@inproceedings{takahashi2026mmaudiosep,
  title={{MMAudioSep}: Taming Video-to-Audio Generative Model Towards Video/Text-Queried Sound Separation},
  author={Takahashi, Akira and Takahashi, Shusuke and Mitsufuji, Yuki},
  booktitle=ICASSP,
  year={2026}
}

@inproceedings{cheng2025omnisep,
  title={{OmniSep}: Unified Omni-Modality Sound Separation with Query-Mixup},
  author={Cheng, Xize and Zheng, Siqi and Wang, Zehan and Fang, Minghui and Zhang, Ziang and Huang, Rongjie and Ji, Shengpeng and Zuo, Jialong and Jin, Tao and Zhao, Zhou},
  booktitle=ICLR,
  year={2025}
}

@article{melechovsky2025sonicmaster,
  title={Sonicmaster: Towards controllable all-in-one music restoration and mastering},
  author={Melechovsky, Jan and Mehrish, Ambuj and Roy, Abhinaba and Herremans, Dorien},
  journal={arXiv preprint arXiv:2508.03448},
  year={2025}
}

@article{roy2026merit,
  title={MERIT: Learning Disentangled Music Representations for Audio Similarity},
  author={Roy, Abhinaba and Liang, Junyi and Herremans, Dorien},
  journal={arXiv preprint arXiv:2605.27346},
  year={2026}
}

@article{roy2018discriminative,
  title={Discriminative body part interaction mining for mid-level action representation and classification},
  author={Roy, Abhinaba and Banerjee, Biplab and Murino, Vittorio},
  journal={Journal of Visual Communication and Image Representation},
  volume={55},
  pages={829--840},
  year={2018},
  publisher={Elsevier}
}
}

% ===========================================================================
% Supplementary material
% ===========================================================================
\clearpage
\maketitlesupplementary
\renewcommand{\thesection}{\Alph{section}}
\renewcommand{\thetable}{S\arabic{table}}
\renewcommand{\theequation}{S\arabic{equation}}
% unique hyperref anchors for the restarted counters
\def\theHsection{supp.\Alph{section}}
\def\theHtable{supp.\arabic{table}}
\def\theHequation{supp.\arabic{equation}}
\setcounter{section}{0}
\setcounter{table}{0}
\setcounter{equation}{0}

This document gives the settings, the full tables and the protocols that the paper refers to.
All scores are median SDR in dB, as in the paper, unless we say otherwise.
\emph{Ours} is the model of the paper, trained with all kinds of mixture from the first step.
The curriculum, no-video, keypoints-only and no-mirror models, and the model without Mega's weights, are ablations (paper, Sec.~4).

\section{Training and inference settings}
\label{sec:supp-train}

\paragraph{Loss.}
The loss of the paper (Sec.~3.3) adds an $\ell_1$ waveform term and five complex-spectrogram $\ell_1$ terms, all with weight 1:
\begin{align}
  \mathcal{L} &= \mathcal{L}_{\mathrm{wav}} + \sum_{r=1}^{5} \mathcal{L}_{\mathrm{spec}}^{(r)} , \\
  \mathcal{L}_{\mathrm{wav}} &= \lVert \hat{s}_a - s_a \rVert_1 , \\
  \mathcal{L}_{\mathrm{spec}}^{(r)} &= \lVert \mathrm{STFT}_r(\hat{s}_a) - \mathrm{STFT}_r(s_a) \rVert_1 ,
\end{align}
where $s_a$ is the target audio of a training mixture and $\hat{s}_a$ the model's estimate of it (paper, Eq.~4).
Here $\lVert \cdot \rVert_1$ is the mean absolute value over all elements, and for the spectrogram terms it covers the real and the imaginary parts.
$\mathrm{STFT}_r$ uses a Hann window of 4096, 2048, 1024, 512 or 256 samples, a hop of 147 samples and no normalisation.
These are the loss settings of the training configuration of MVSep Mega~\cite{solovyev2026musicsourceseparationtrainingmsstunifiedframework}, from which our backbone starts.

\paragraph{Optimisation.}
We train with AdamW (learning rate $3 \times 10^{-4}$, $\beta = (0.9, 0.95)$, weight decay 0.01), 1,000 warm-up steps and a cosine schedule over 52k steps.
The pretrained backbone uses one sixth of the learning rate of the new layers.
A batch holds 32 mixtures of 4\,s, eight on each of four GPUs.
For every mixture the second player is drawn as a different instrument (probability 0.20), the same instrument from another video (0.30), the same instrument from another video matched in pitch and tempo (0.25), or the same performer at another moment (0.25), and zero to three clips of other instruments are added.
We measure the SNR of the target player on a fixed set of validation mixtures, on which the unprocessed mixture scores \m3.97\dB, and keep the checkpoint with the highest SNR.
Our model reaches its best validation SNR, 10.27\dB, at step 37k.
Training took 13.7 hours on four RTX PRO 6000 GPUs.
The V-JEPA~2 tokens of the mirrored videos are computed once before training.

\paragraph{Model size.}
Our model has 43.9M trained parameters, of which 26.3M are the pretrained backbone.
The keypoints-only model has 43.6M trained parameters.
SAM~2.1 (Hiera-L), V-JEPA~2 (ViT-L) and RTMW are frozen and not counted.

\paragraph{Inference.}
We process a piece in windows of 4\,s with 50\% overlap and blend the windows with a Hann window.
The pretrained backbone expects stereo input, so we feed the mono recording on both channels and average the two outputs.

\section{How the baselines were run}
\label{sec:supp-baselines}

No baseline was trained or tuned by us.

\paragraph{SAM Audio.}
We use the released large checkpoint, \texttt{sam-audio-large-tv}.
The visual prompt is the SAM~2.1 mask of the clicked player.
The rankers are switched off, and a piece is processed in chunks of 10\,s with 1\,s overlap.

\paragraph{iQuery.}
We use the authors' code and their weights trained on MUSIC.
The model receives the crop of the clicked player and the name of the player's instrument as its query.

\paragraph{DAVIS-Flow.}
We use the authors' code and their weights trained on MUSIC.
The model receives the crop of the clicked player.
It is trained to return one source from the average of two sources, so we scale its output by one half.
Without this scaling it scores \m2.40\dB over all players and 0.01\dB on pure duets, against 0.01 and 0.02\dB with it.

\paragraph{MVSep Mega.}
The model is told the instrument of the clicked player and returns the stem of that instrument, which holds every player of it.

\paragraph{Sampling rate.}
iQuery and DAVIS-Flow output audio at 11\,kHz.
All outputs are resampled to 48\,kHz and scored against the full-band track, so these two models lose the content above 5.5\,kHz.

\paragraph{Held-out mixtures.}
iQuery is scored on the 220 mixtures whose instrument is one of its 11 classes, and Mega on the 408 mixtures for which it has a stem.
A solo video shows one person, so SAM Audio receives the clip without a mask, and iQuery and DAVIS-Flow receive the whole frame.
255 of the 500 mixtures use a video that is in the MUSIC training lists of iQuery or DAVIS-Flow.
Without these mixtures our model reaches 9.71\dB over all mixtures and 13.54\dB for two players only, and every baseline stays at or below 0.57\dB.

\paragraph{MUSIC benchmark.}
To check that we run iQuery and DAVIS-Flow correctly, we score them on the benchmark of their own papers (paper, Tab.~3).
We build 200 mixtures of two different instruments from 77 test recordings of MUSIC and MUSIC-21, which cover the 11 MUSIC instruments.
None of these recordings was used to train our model or to pick its checkpoint.
We follow the protocol of the two baselines.
A window is 5.94\,s long (65,535 samples).
Each model receives the mixture and whole frames exactly as in its own code: iQuery and our model receive the sum of the two sources, and DAVIS-Flow their average.
The score is the BSS Eval v3 SDR over the window, averaged over both sources and all mixtures, at 11,025\,Hz, to which our output is resampled.
Both baselines come within 0.5\dB of their published scores: iQuery reaches 11.61\dB (11.17 in its paper) and DAVIS-Flow 12.21\dB (12.01 in its paper).
Our model reaches 19.08\dB and the curriculum model 18.89\dB.
Our model is better on 183 of the 200 mixtures against DAVIS-Flow and on 187 against iQuery.

\paragraph{Methods we could not run.}
We found no public code for Sound of Motions~\cite{zhao2019sound} and Music Gesture~\cite{gan2020music}, and we could not obtain weights for Sound of Pixels~\cite{zhao2018sound}, MMAudioSep~\cite{takahashi2026mmaudiosep} and OmniSep~\cite{cheng2025omnisep}.

\section{Intervals and counts}
\label{sec:supp-stats}

\begin{table}[t]
  \centering
  \footnotesize
  \setlength{\tabcolsep}{3.4pt}
  \begin{tabular}{@{}lcc@{}}
    \toprule
    Row & Median [95\% interval] & Ours better \\
    \midrule
    Unchanged recording        & \m6.29 [\m6.79, \m5.61] & 58 of 58 \\
    Scaled recording           & 0.82 [0.66, 1.00]       & 46 of 58 \\
    SAM Audio                  & \m3.92 [\m4.58, \m3.16] & 57 of 58 \\
    iQuery                     & \m0.45 [\m0.78, \m0.10] & 51 of 58 \\
    DAVIS-Flow                 & 0.01 [\m0.09, 0.01]     & 47 of 58 \\
    MVSep Mega                 & 0.30 [0.19, 0.78]       & 42 of 58 \\
    \midrule
    Curriculum model           & 3.43 [1.01, 6.79]       & 44 of 58 \\
    Ours, still frame          & \m0.97 [\m1.69, \m0.47] & 50 of 58 \\
    \textbf{Ours}              & 5.30 [1.86, 10.31]      &          \\
    \bottomrule
  \end{tabular}
  \caption{\textbf{Intervals and counts over all 58 players.}
  \emph{Ours better}: players for whom our model scores higher than the row.}
  \label{tab:supp-stats}
\end{table}

\cref{tab:supp-stats} gives the interval and the count for every row, including the curriculum model and the still-frame test of the paper.
The interval of the median comes from 10,000 bootstrap resamples in which the two players of a pair are drawn together.

\paragraph{Pure duets.}
For the eight players of the four pure duets the interval of our model is 11.74 to 20.16\dB (median 18.73).
Our model is ahead of every released baseline for all eight players, and of the scaled recording (2.91\dB) as well.
These intervals rest on four pieces and are rough.

\section{Results by ensemble size and by instrument}
\label{sec:supp-groups}

\begin{table*}[t]
  \centering
  \footnotesize
  \setlength{\tabcolsep}{4.5pt}
  \begin{tabular}{@{}lcccccccccc@{}}
    \toprule
    & & Unchanged & Scaled & & & & MVSep & Curriculum & Keypoints & \\
    Group & $n$ & recording & recording & SAM Audio & iQuery & DAVIS-Flow & Mega & model & only & \textbf{Ours} \\
    \midrule
    \multicolumn{11}{@{}l}{\emph{Players in the piece}} \\
    2 & 8  & 0.00   & 2.91 & \m0.20 & 0.29   & 0.02   & 1.75 & 15.66  & 12.18 & 18.73 \\
    3 & 10 & \m4.56 & 1.18 & \m2.88 & \m0.52 & 0.00   & 0.80 & 2.03   & 1.08  & 1.80 \\
    4 & 24 & \m6.70 & 0.68 & \m4.32 & \m0.57 & 0.01   & 0.38 & 6.07   & 6.03  & 10.34 \\
    5 & 16 & \m6.95 & 0.72 & \m4.90 & \m0.13 & \m0.29 & 0.26 & \m0.06 & 0.26  & 1.14 \\
    \midrule
    \multicolumn{11}{@{}l}{\emph{Instrument of the clicked player}} \\
    Violin    & 26 & & 0.68 & \m4.06 & \m0.54 & 0.01   & 0.48   & 3.66  & 5.31  & 6.88 \\
    Viola     & 2  & & 0.94 & \m2.68 & \m1.80 & \m1.84 & \m0.16 & 1.63  & 1.74  & 4.03 \\
    Flute     & 12 & & 1.13 & \m3.16 & 0.00   & 0.02   & 0.36   & 7.05  & 4.23  & 8.56 \\
    Trumpet   & 16 & & 0.74 & \m4.40 & \m0.75 & 0.00   & 0.18   & 0.68  & 1.08  & 1.14 \\
    Saxophone & 2  & & 2.84 & 4.30   & 0.01   & 0.28   & 1.76   & 16.77 & 15.34 & 14.78 \\
    \bottomrule
  \end{tabular}
  \caption{\textbf{URMP by ensemble size and by instrument.}
  $n$ is the number of clicked players.
  The curriculum and keypoints-only models are ablations.
  Viola and saxophone are one pair each, which is too few to draw a conclusion for these instruments.}
  \label{tab:supp-groups}
\end{table*}

\cref{tab:supp-groups} holds the numbers behind Fig.~5 of the paper, for every method.
Our model is above the scaled recording in every group, by the least in three-player pieces (1.80 against 1.18\dB), in five-player pieces (1.14 against 0.72\dB) and for trumpets (1.14 against 0.74\dB).

\section{Held-out mixtures}
\label{sec:supp-heldout}

\begin{table}[t]
  \centering
  \footnotesize
  \setlength{\tabcolsep}{3.3pt}
  \begin{tabular}{@{}lccccc@{}}
    \toprule
    Players in the mixture & 2 & 3 & 4 & 5 & All \\
    \midrule
    $n$                                 & 134    & 126    & 121    & 119    & 500 \\
    Unchanged mixture                   & \m0.26 & \m4.19 & \m5.84 & \m7.45 & \m4.46 \\
    Scaled mixture, $x/N$               & 2.65   & 1.28   & 0.94   & 0.62   & 1.09 \\
    SAM Audio                           & \m0.07 & \m2.19 & \m3.31 & \m3.85 & \m2.47 \\
    iQuery$^{\ast}$                     & 0.46   & \m1.03 & \m0.46 & \m2.25 & \m0.66 \\
    DAVIS-Flow                          & 0.29   & \m0.04 & \m0.23 & \m0.97 & \m0.18 \\
    MVSep Mega$^{\dagger}$              & 0.14   & 0.80   & 1.08   & \m0.04 & 0.26 \\
    \midrule
    Keypoints only                      & 11.68  & 8.43   & 5.34   & 3.58   & 6.46 \\
    Curriculum model                    & 14.31  & 10.82  & 8.93   & 5.85   & 9.90 \\
    \textbf{Ours}                       & 14.36  & 10.56  & 8.42   & 6.21   & 9.56 \\
    \bottomrule
  \end{tabular}
  \caption{\textbf{Held-out mixtures by the number of players.}
  Each mixture holds two players of one instrument and zero to three other instruments.
  $^{\ast}$Only mixtures whose instrument is one of iQuery's 11 classes: 63, 54, 59 and 44, 220 in all.
  $^{\dagger}$Only mixtures for which Mega has a stem: 114, 107, 91 and 96, 408 in all.
  The column for two players is the held-out \emph{pure duets only} column of paper Tab.~1.}
  \label{tab:supp-heldout-size}
\end{table}

\begin{table}[t]
  \centering
  \footnotesize
  \setlength{\tabcolsep}{2.4pt}
  \begin{tabular}{@{}lccccc@{}}
    \toprule
    & & No & Keypoints & Curriculum & \\
    Second player & $n$ & video & only & model & \textbf{Ours} \\
    \midrule
    Another video                & 170 & 1.22 & 8.49  & 11.46 & 10.87 \\
    Another video, matched       & 149 & 1.18 & 8.06  & 10.47 & 10.63 \\
    Same video, another moment   & 181 & 1.39 & 3.86  & 6.51  & 6.31 \\
    \midrule
    All mixtures                 & 500 & 1.24 & 6.46  & 9.90  & 9.56 \\
    Two players only             & 134 & 2.44 & 11.68 & 14.31 & 14.36 \\
    \bottomrule
  \end{tabular}
  \caption{\textbf{Held-out mixtures by the kind of second player.}
  \emph{Matched}: the other video is matched in pitch and tempo.
  The hardest case is the same performer at another moment, where the two players look identical.
  \emph{Two players only}: the held-out \emph{pure duets only} column of paper Tab.~1.}
  \label{tab:supp-heldout-kind}
\end{table}

\cref{tab:supp-heldout-size} splits the 500 held-out mixtures by the number of players.
No baseline reaches 1.1\dB for any number of players, and only Mega with four players (1.08\dB) is above the scaled mixture (0.94\dB).
Every model drops as players are added, and our model goes from 14.36\dB with two players to 6.21\dB with five.
\cref{tab:supp-heldout-kind} splits the same mixtures by how the second player was chosen.
In every kind the order of the ablations is the same: no video, then keypoints only, then the curriculum model.
Our model and the curriculum model are within 0.6\dB of each other in every kind.
These mixtures are built like our training mixtures, so they are in domain for our models and out of domain for the baselines.

\section{Time shift}
\label{sec:supp-shift}

\begin{table}[t]
  \centering
  \footnotesize
  \setlength{\tabcolsep}{3.2pt}
  \begin{tabular}{@{}lcccccc@{}}
    \toprule
    & \multicolumn{2}{c}{\textbf{Ours}} & \multicolumn{2}{c}{Curriculum} & \multicolumn{2}{c}{Keypoints only} \\
    \cmidrule(lr){2-3} \cmidrule(lr){4-5} \cmidrule(l){6-7}
    Shift (s) & All & Duets & All & Duets & All & Duets \\
    \midrule
    0     & 5.30   & 18.73 & 3.43   & 15.66 & 4.15   & 12.18 \\
    0.04  &        &       &        &       & 2.36   & 6.90 \\
    0.08  & 1.38   & 10.54 & 0.67   & 9.14  & 0.45   & 2.93 \\
    0.12  &        &       &        &       & 0.17   & 2.17 \\
    0.16  & 0.11   & 6.79  & \m0.05 & 6.90  & \m0.04 & 3.89 \\
    0.20  &        &       &        &       & \m0.35 & 3.47 \\
    0.24  & \m0.06 & 8.00  & \m0.63 & 8.90  &        &      \\
    0.4   & \m0.32 & 8.82  & \m0.98 & 2.50  & \m0.68 & 1.67 \\
    1.2   & \m0.57 & 4.73  & \m0.68 & 0.67  & \m0.67 & 1.48 \\
    2     & \m0.23 & 5.66  & \m0.61 & 2.08  & \m0.41 & 2.54 \\
    4     & \m0.54 & 6.43  & \m0.83 & 4.37  & \m0.71 & 1.02 \\
    \midrule
    No-video model & 0.52 & 2.95 & 0.52 & 2.95 & 0.52 & 2.95 \\
    \bottomrule
  \end{tabular}
  \caption{\textbf{Time shift.}
  Median SDR in dB over all 58 players and over the 8 players in pure duets, when the player's video is shifted later than the audio by the given amount.
  Empty cells were not run.
  Models with V-JEPA~2 tokens can only be shifted in steps of 0.08\,s, the time that one token covers.}
  \label{tab:supp-shift}
\end{table}

For every clicked player we shift the player tokens later than the audio by a fixed amount and wrap them around at the end of the piece.
For models with V-JEPA~2 tokens, these tokens and the keypoint tokens are shifted together.
All models use the checkpoints of the paper and nothing is retrained.
\cref{tab:supp-shift} gives every shift we ran.

\paragraph{Player by player.}
At a shift of 0.08\,s, 45 of the 58 players score lower than without a shift with our model, 46 with the curriculum model and 44 with the keypoints-only model.
The median drop per player is 1.06, 1.04 and 1.64\dB.
For our model the count rises to 46 at 0.16\,s, 48 at 0.4\,s and 50 at 4\,s.

\section{Amount of movement}
\label{sec:supp-movement}

\begin{table}[t]
  \centering
  \footnotesize
  \setlength{\tabcolsep}{4pt}
  \begin{tabular}{@{}lcccc@{}}
    \toprule
    & & & & Arm speed, \\
    Players & $n$ & Arm speed & Hand speed & loudness removed \\
    \midrule
    All     & 58 & 0.07 (0.60)   & 0.03 (0.80)   & 0.18 (0.17) \\
    Violin  & 26 & 0.11 (0.61)   & 0.31 (0.13)   & 0.24 (0.24) \\
    Trumpet & 16 & 0.19 (0.47)   & 0.12 (0.66)   & 0.19 (0.47) \\
    Flute   & 12 & \m0.23 (0.47) & \m0.43 (0.17) & \m0.17 (0.60) \\
    \bottomrule
  \end{tabular}
  \caption{\textbf{Does a player who moves more gain more?}
  Spearman correlation between a player's mean keypoint speed and the gain of our model over the no-video model, with the $p$-value in brackets.}
  \label{tab:supp-movement}
\end{table}

For every clicked player we take the keypoint velocities that the model receives and compute the mean speed over the piece, over the frames in which the keypoint is detected.
We do this for the 6 arm keypoints and for the 42 hand keypoints.
The gain of a player is the score of our model minus the score of the no-video model.
\cref{tab:supp-movement} shows that no correlation is significant, over all players or within an instrument.
Louder players may also move more, so the last column removes each player's loudness in the recording (partial correlation).
Within a pair, the player who moves more gains more in 13 of the 29 pairs.
The gain does correlate with the player's loudness in the recording (Spearman 0.37, $p = 0.004$).
The curriculum model gives the same picture: over all players the three correlations are 0.04, \m0.05 and 0.16, and the player who moves more gains more in 12 of the 29 pairs.
We tested mean speed only and did not search for other measures of movement.

\section{Click test}
\label{sec:supp-click}

\begin{table}[t]
  \centering
  \footnotesize
  \setlength{\tabcolsep}{2.6pt}
  \begin{tabular}{@{}lcccc@{}}
    \toprule
    & \multicolumn{2}{c}{All (58)} & \multicolumn{2}{c}{Pure duets (8)} \\
    \cmidrule(lr){2-3} \cmidrule(l){4-5}
    Method & correct / other & wins & correct / other & wins \\
    \midrule
    SAM Audio                 & \m3.92 / \m3.98 & 29 & \m0.20 / 0.20  & 4 \\
    iQuery                    & \m0.45 / \m0.34 & 23 & 0.29 / 0.69    & 1 \\
    DAVIS-Flow                & 0.01 / 0.01     & 27 & 0.02 / 0.01    & 4 \\
    \midrule
    (b) curriculum            &                 & 52 &                & 8 \\
    (d) keypoints only        &                 & 51 &                & 8 \\
    (e) no mirror augm.       &                 & 47 &                & 8 \\
    (g) not pretrained        &                 & 49 &                & 5 \\
    (g), full resolution      &                 & 48 &                &   \\
    Ours, still frame         &                 & 26 &                &   \\
    \midrule
    \textbf{Ours}             & 5.30 / \m2.19   & 51 & 18.73 / \m1.76 & 8 \\
    \bottomrule
  \end{tabular}
  \caption{\textbf{Click test.}
  Median SDR with the click on the correct player and on the other player of the pair, and the number of players for whom the correct click scores higher (\emph{wins}).
  The rows in the middle are the ablations of paper Tab.~5, with the same letters.
  Row (g) is scored at 960\,px as in the paper, and also at full resolution.
  Empty cells were not reported.
  MVSep Mega and the no-video model return the same output for both clicks, so the correct click never wins.}
  \label{tab:supp-click}
\end{table}

\cref{tab:supp-click} gives the click test for all models.
For our model the correct click wins for 24 of the 28 string players and for 27 of the 30 wind and brass players.

\section{Leakage split}
\label{sec:supp-leak}

The paper (Tab.~2) splits the unwanted sound in each output into the partner and the other instruments.
We cut the output $y$ into windows of 1\,s at 48\,kHz and fit each window by least squares,
\begin{equation}
  y \approx a\,s + b\,q + c\,r ,
\end{equation}
where $s$ is the clicked player's true track, $q$ the partner's true track, and $r$ the recording minus $s$ and $q$, which holds the other instruments and is zero in pure duets.
The two leakage values of a window are
\begin{equation}
  10 \log_{10} \frac{\lVert a s \rVert^2}{\lVert b q \rVert^2}
  \quad \text{and} \quad
  10 \log_{10} \frac{\lVert a s \rVert^2}{\lVert c r \rVert^2} ,
\end{equation}
for the partner and for the other instruments.
Windows in which the clicked player is silent are skipped.
We take the median over windows and then over players.
This fit uses one gain per signal and window, not the 512-tap filters of BSS Eval, so its values are not the SIR of museval.

\section{Other prompts}
\label{sec:supp-prompts}

\paragraph{Click variation.}
Each of the 58 players is prompted three more ways in place of the three default points (nose, shoulders, hips): the nose alone, a box around the player on the first frame, and the three points each moved at random by up to 10\% of the player's height.
Each prompt goes through SAM~2.1 tracking, keypoints, full-resolution V-JEPA~2 features and our model, with no retraining.
More than 1\dB worse than the default are 8 players with the nose alone, 5 with the box and 7 with the moved points.
More than 1\dB better are 2, 4 and 2 players.
The worst single case is one violinist who scores 9.20\dB with the default prompt and 1.13\dB with the nose alone.
The curriculum model gives 3.22, 3.04 and 2.95\dB over all players with the nose alone, the box and the moved points, against 4.25, 4.64 and 5.63\dB for our model.

\paragraph{Text prompt.}
SAM~3 receives a phrase such as ``left violin'' and returns a mask, which replaces the click.
The separator is unchanged.
SAM~3 selects the intended player for 33 of the 58 players and for all eight players in pure duets.
It selects another player for 11 players, for example when it takes a viola for a violin, and the sound of the wrong player then scores between \m4.6 and \m0.6\dB.
It finds no matching player for 14 players, most of them flutes.
For these nothing is separated, and we score the unchanged recording.
When SAM~3 selects the intended player, the output is the same as with the click.

\paragraph{A click on nobody.}
For each of the 29 pieces we place one click on the background of the first frame, at the point of a grid with a spacing of 5\% of the frame that is farthest from every detected person.
These points are 240 to 320\,px away from the nearest person.
The click goes through the full pipeline.
The median level of the output is \m7.5\dB against the recording, with a range from \m19.4 to \m4.0\dB, against \m6.1\dB for a real click.
The output is quieter in pieces with winds and brass (\m10 to \m19\dB in trumpet pieces) and close to the level of a real click in string pieces.
Against the closer of the two players of the pair it scores a median SDR of \m0.07\dB, against 5.30\dB for a real click, so it returns a part of the music that belongs to no one player.

\section{Curriculum ablation}
\label{sec:supp-curriculum}

\paragraph{Schedule.}
The curriculum model differs from ours only in the order of the training mixtures.
It trains for 4k steps with a different instrument as the second player, then 4k steps with the same instrument from another video, then 4k steps with the same performer at another moment, and from step 12k on with the full mixtures of our model.
The model, the data, the mirror augmentation, the step budget, the learning-rate schedule and the checkpoint selection are the same.
Its validation SNR is 0.99, 3.10 and 4.29\dB at the end of the three stages and 9.81\dB at its best checkpoint, step 44k.
Early stopping, after five validation checks without improvement, ended the run at step 49k, after 17.6 hours on four RTX PRO 6000 GPUs.

\begin{table}[t]
  \centering
  \footnotesize
  \setlength{\tabcolsep}{4pt}
  \begin{tabular}{@{}lcccc@{}}
    \toprule
    Group & $n$ & \textbf{Ours} & Curriculum & Ours higher for \\
    \midrule
    2 players (pure duets) & 8  & \textbf{18.73} & 15.66 & 6 of 8 \\
    3 players              & 10 & 1.80 & \textbf{2.03}   & 8 of 10 \\
    4 players              & 24 & \textbf{10.34} & 6.07  & 20 of 24 \\
    5 players              & 16 & \textbf{1.14} & \m0.06 & 10 of 16 \\
    \midrule
    Violin                 & 26 & \textbf{6.88} & 3.66   & 19 of 26 \\
    Viola                  & 2  & \textbf{4.03} & 1.63   & 2 of 2 \\
    Flute                  & 12 & \textbf{8.56} & 7.05   & 10 of 12 \\
    Trumpet                & 16 & \textbf{1.14} & 0.68   & 13 of 16 \\
    Saxophone              & 2  & 14.78 & \textbf{16.77} & 0 of 2 \\
    \midrule
    All                    & 58 & \textbf{5.30} & 3.43   & 44 of 58 \\
    \bottomrule
  \end{tabular}
  \caption{\textbf{Regular training (ours) against the curriculum on URMP.}
  Median SDR in dB.}
  \label{tab:supp-curr-urmp}
\end{table}

\begin{table}[t]
  \centering
  \footnotesize
  \setlength{\tabcolsep}{4pt}
  \begin{tabular}{@{}lcc@{}}
    \toprule
    Test & \textbf{Ours} & Curriculum \\
    \midrule
    Held-out, 2 players             & 14.36 & 14.31 \\
    Held-out, 3 players             & 10.56 & 10.82 \\
    Held-out, 4 players             & 8.42 & 8.93 \\
    Held-out, 5 players             & 6.21 & 5.85 \\
    Held-out, all 500               & 9.56 & 9.90 \\
    Held-out, another video         & 10.87 & 11.46 \\
    Held-out, matched               & 10.63 & 10.47 \\
    Held-out, same performer        & 6.31 & 6.51 \\
    \midrule
    Correct click wins, all / duets & 51 / 8 & 52 / 8 \\
    MUSIC benchmark                 & 19.08 & 18.89 \\
    Nose / box / moved points, all  & 4.25 / 4.64 / 5.63 & 3.22 / 3.04 / 2.95 \\
    Time shift 0.08\,s, all / duets & 1.38 / 10.54 & 0.67 / 9.14 \\
    Time shift 0.4\,s, all / duets  & \m0.32 / 8.82 & \m0.98 / 2.50 \\
    Validation SNR                  & 10.27 & 9.81 \\
    \bottomrule
  \end{tabular}
  \caption{\textbf{Regular training (ours) against the curriculum on the other tests.}
  On the held-out mixtures our model is higher on 271 of the 500 mixtures.}
  \label{tab:supp-curr-other}
\end{table}

\paragraph{Results.}
\cref{tab:supp-curr-urmp,tab:supp-curr-other} compare the two models.
Our model is better on URMP over all players, on pure duets, in four- and five-player pieces, for violin, viola, flute and trumpet, on the validation mixtures, on the MUSIC benchmark and with every other prompt.
The curriculum model has the higher median in three-player pieces, although our model is higher for 8 of those 10 players, and on the single saxophone pair.
On the held-out mixtures the two do not differ by much: 9.56 against 9.90\dB, and our model is higher on 271 of the 500 mixtures.
The curriculum model is higher with three and four players and for two of the three kinds of second player, by 0.2 to 0.6\dB, and our model is higher with two and five players and for matched videos.
On pure duets the curriculum model loses more of its score when the video is shifted by 0.4\,s or more (\cref{tab:supp-shift}).
We trained each model once, so we do not claim that the curriculum causes this.

\section{Failure cases}
\label{sec:supp-failures}

Our model is below the scaled recording for 12 of the 58 players and does not beat MVSep Mega for 16 (paper, Tab.~1).

\paragraph{Trumpets next to other brass.}
Five trumpet pairs play next to a trombone, a horn or a tuba.
With our model the two players of each pair score 1.07 and \m3.22, 1.21 and \m0.85, 0.37 and \m2.00, 2.55 and \m0.50, and \m3.41 and \m2.06\dB.
The trumpet duet scores 19.73 and 20.58\dB.

\paragraph{Two violin pairs.}
Two violin pairs in five-player pieces fail with our model.
The four players score \m3.32, \m2.30, \m3.12 and \m2.24\dB, all below the scaled recording.
The tracking is correct and the two parts of each pair do not play the same notes.
We do not know why these pairs fail.

\paragraph{Worst players.}
The four lowest scores are \m3.41\dB (a trumpet), \m3.32\dB (a violin), \m3.22\dB (a trumpet) and \m3.12\dB (a violin).

\section{Listening test protocol}
\label{sec:supp-listening}

This section gives the protocol of the listening test in Sec.~5.2 and Tab.~4 of the paper.
The test follows MUSHRA~\cite{schoeffler2015towards} and runs in a web browser.

\begin{table}[t]
  \centering
  \footnotesize
  \setlength{\tabcolsep}{5pt}
  \begin{tabular}{@{}ccccc@{}}
    \toprule
    Trial & URMP piece & Players & Clicked player & Start (s) \\
    \midrule
    1  & 04 & 2 & flute   & 5  \\
    2  & 05 & 2 & trumpet & 29 \\
    3  & 02 & 2 & violin  & 5  \\
    4  & 14 & 3 & flute   & 15 \\
    5  & 13 & 3 & violin  & 5  \\
    6  & 25 & 4 & violin  & 29 \\
    7  & 34 & 4 & trumpet & 83 \\
    8  & 29 & 4 & flute   & 11 \\
    9  & 40 & 5 & flute   & 5  \\
    10 & 44 & 5 & viola   & 26 \\
    \bottomrule
  \end{tabular}
  \caption{\textbf{Listening-test trials.}
  One trial per URMP piece. \emph{Players}: players in the piece. \emph{Start}: where the 10\,s excerpt begins.
  Trials 1 to 3 are the pure duets.}
  \label{tab:supp-listening}
\end{table}

\paragraph{Trial.}
A trial shows a still frame of the video with the clicked player in a red box.
The page tells the listener that the picture only shows which player was clicked, and that the clips are to be judged by ear.
The listener can play the reference, which is the clicked player's true track, and seven unlabelled clips: the hidden reference, our model, MVSep Mega, SAM Audio, iQuery, DAVIS-Flow, and the unchanged recording as the anchor.
The listener rates each clip from 0 to 100 for how close it sounds to the reference, on a scale labelled bad, poor, fair, good and excellent.
The instructions ask the listener to mark a clip down for every other player they can hear and for distortion or missing notes, and they say that one of the clips is the reference itself.
A slider unlocks only after its clip has been played, and every clip can be replayed as often as the listener likes.
The scaled recording is left out, because at matched loudness it sounds the same as the anchor.

\paragraph{Trials and clips.}
There are 10 trials, one per URMP piece (\cref{tab:supp-listening}): 3 pure duets, 2 trios, 3 quartets and 2 quintets, with 4 string and 6 wind or brass players.
The trials are drawn at random with a fixed seed, balanced over these groups, and no piece appears twice.
The excerpt of a trial is the first 10\,s window, after the first 5\,s of the piece, in which both players of the pair play for at least 9\,s.
A player counts as playing when their true track is above \m40\,dBFS.
This rule was fixed before anyone listened to an output.
Every clip is cut from the full-piece output that is scored in the paper, normalised to \m23\,LUFS and stored as a mono WAV file at 44.1\,kHz.
The order of the trials, and of the clips within a trial, is shuffled for every listener.

\paragraph{Listeners.}
16 listeners took part, and none of them is an author of this paper.
The page asks them to wear headphones in a quiet room.
Progress is saved in the browser, so a listener can stop and continue later.
No listener was removed.

\paragraph{Analysis.}
For each method we report the mean score with a 95\% bootstrap interval, resampled over listeners and trials, over all 10 trials and over the 3 duet trials (paper, Tab.~4).
To compare our model with each baseline, we take each listener's mean rating of each method and apply an exact two-sided Wilcoxon signed-rank test over the 16 listeners, with Holm correction over the four baselines.
Every listener gives our model a higher mean rating than any baseline, so each test gives $p = 2/2^{16} \approx 3.1 \times 10^{-5}$, which is $1.2 \times 10^{-4}$ after correction.

\end{document}